\documentclass[twocolumn,longauthor,longbib]{aastex7}
\usepackage{amsmath}

\newcommand{\spr}{S^\prime}
\newcommand{\AAng}{\text{\AA}}

\begin{document}

\title{Post-fragmentation Morphological and Broadband Color Evolution of C/2025 K1 (ATLAS)}
\shorttitle{Post-fragmentation Evolution of C/2025 K1}

\author[0009-0000-0948-1320,gname={Kody M.},sname={Walker}]{Kody M. Walker}

\affiliation{Department of Physics and Astronomy, East Texas A\&M University, Commerce, TX 75428, USA}

\email[show]{kwalker82@leomail.etamu.edu}

\begin{abstract}
Comet C/2025 K1 (ATLAS) underwent major fragmentation after its 2025 October 8 perihelion passage at $q\simeq0.33$ au. We present four epochs of Johnson--Cousins $BVR$ imaging obtained with the Mount Laguna Observatory 40-inch reflector on 2025 November 26, November 29, December 5, and December 8, sampling the subsequent evolution of the integrated fragment-coma complex. The images were homogenized to common circular two-dimensional effective point-spread functions before morphology was measured along the projected fragment axis. We evaluated 324 profile-extraction configurations and removed solutions whose localized peak landed on the search-window boundary. The fragment-axis surface-brightness enhancement relocates inward between November 26 and 29 and outward between December 5 and 8; the sign of both changes is recovered in every valid configuration and at all tested common target PSFs from 5.3 to 6.0 arcsec. Calibrated $BVR$ photometry using Gaia DR3 synthetic Johnson--Kron--Cousins standards independently shows a repeatable November 26--29 change in broadband spectral shape: $B-V$ decreases while $V-R$ increases. The curvature index $\spr_{V-R}-\spr_{B-V}$ changes by $+3.47^{+1.50}_{-1.65}$\% per 1000~\AAng, with 99.1\% of Monte Carlo realizations positive, while the full $B-R$ reflectivity gradient remains approximately constant. Repeating the color comparison at a fixed projected radius of 2205 km produces nearly identical changes. The combined results are consistent with continued restructuring of the unresolved post-fragmentation inner coma, while broadband imaging alone cannot uniquely separate dust-color evolution, molecular emission, and changing contributions from unresolved fragments.
\end{abstract}


\section{Introduction}\label{sec:intro}

Fragmentation is an important pathway in the physical evolution and eventual destruction of cometary nuclei. Split comets can produce multiple active components whose individual comae and tails overlap, whose relative brightnesses evolve with time, and whose fragments may themselves undergo further disruption \citep{Boehnhardt2004}. The resulting post-fragmentation morphology can therefore represent a time-dependent superposition of several active components rather than a simple collection of isolated point sources. A variety of mechanisms may contribute to cometary disruption. In particular, anisotropic outgassing can exert torques capable of driving small active nuclei toward rotational instability, and rotational breakup has been proposed as an important destruction mechanism for long-period comets with small perihelion distances \citep{Jewitt2022}. Observations obtained after a breakup can consequently provide information not only about the fragmentation event itself but also about the subsequent redistribution of activity, dust, and unresolved fragment contributions.

Broadband optical colors provide a complementary diagnostic of this evolution. Cometary dust continua are commonly redder than the solar spectrum at optical wavelengths, although substantial variation exists among individual objects \citep{JewittMeech1986}. Larger samples likewise show that the mean optical colors of cometary dust are broadly similar between short-period and long-period comets, despite their different dynamical source populations \citep{Jewitt2015}. Temporal changes in broadband color can therefore provide a sensitive measure of changes in the integrated optical spectral shape of a coma. Such measurements do not, however, necessarily trace dust alone. Molecular emission, aperture-dependent coma structure, and changing contributions from unresolved fragments can modify broadband colors, particularly following a major fragmentation event. Broadband measurements are therefore most conservatively interpreted as integrated spectral-shape diagnostics unless the gas and dust contributions can be separated independently.

C/2025 K1 (ATLAS) provides an unusually well-constrained opportunity to examine such post-fragmentation evolution. The comet reached perihelion on 2025 October 8 at approximately $q=0.33$ au and subsequently underwent a major disruption. Ground-based monitoring recorded a strong activity increase during November 2--4, and Hubble Space Telescope (HST) acquisition images obtained on November 8--10 resolved at least five fragments \citep{Bodewits2026}. Back-extrapolation of the fragment motions placed the principal breakup near the beginning of November and showed that the major fragmentation preceded the associated photometric brightening by roughly one to three days. The HST observations further revealed hierarchical fragmentation, including secondary breakup within the fragment system \citep{Bodewits2026}. These observations establish the geometry and timing of the initial disruption, but the system continued to evolve after the earliest resolved HST epochs.

The post-fragmentation optical spectrum introduces an additional complication. Spectroscopy obtained on 2025 November 13 was dominated by NH$_2$ emission, while CN, C$_2$, and C$_3$ were reported as very weak \citep{Ganesh2025}. Johnson--Cousins broadband colors obtained during the following weeks therefore cannot automatically be interpreted as pure dust-continuum colors. Changes in molecular emission, dust scattering properties, and the relative contributions of unresolved fragments and their surrounding comae could all affect the integrated measurements. Later Gemini North and Apache Point Observatory (APO) imaging provides an important endpoint for this period of evolution. \citet{Ingebretsen2026} report Gemini imaging on December 6 and 24 and ARC/APO multiband imaging on December 8, identify at least four distinct fragments, and measure an unusually blue Sloan $g-r$ color near 0.40. Their fragment designations, filter system, angular resolution, and apertures differ from those used here, however, so their numerical colors and component labels should not be mapped directly onto the Horizons B/C/D labels or the Johnson--Cousins measurements in this work.

\begin{figure*}
\centering
\includegraphics[width=\textwidth]{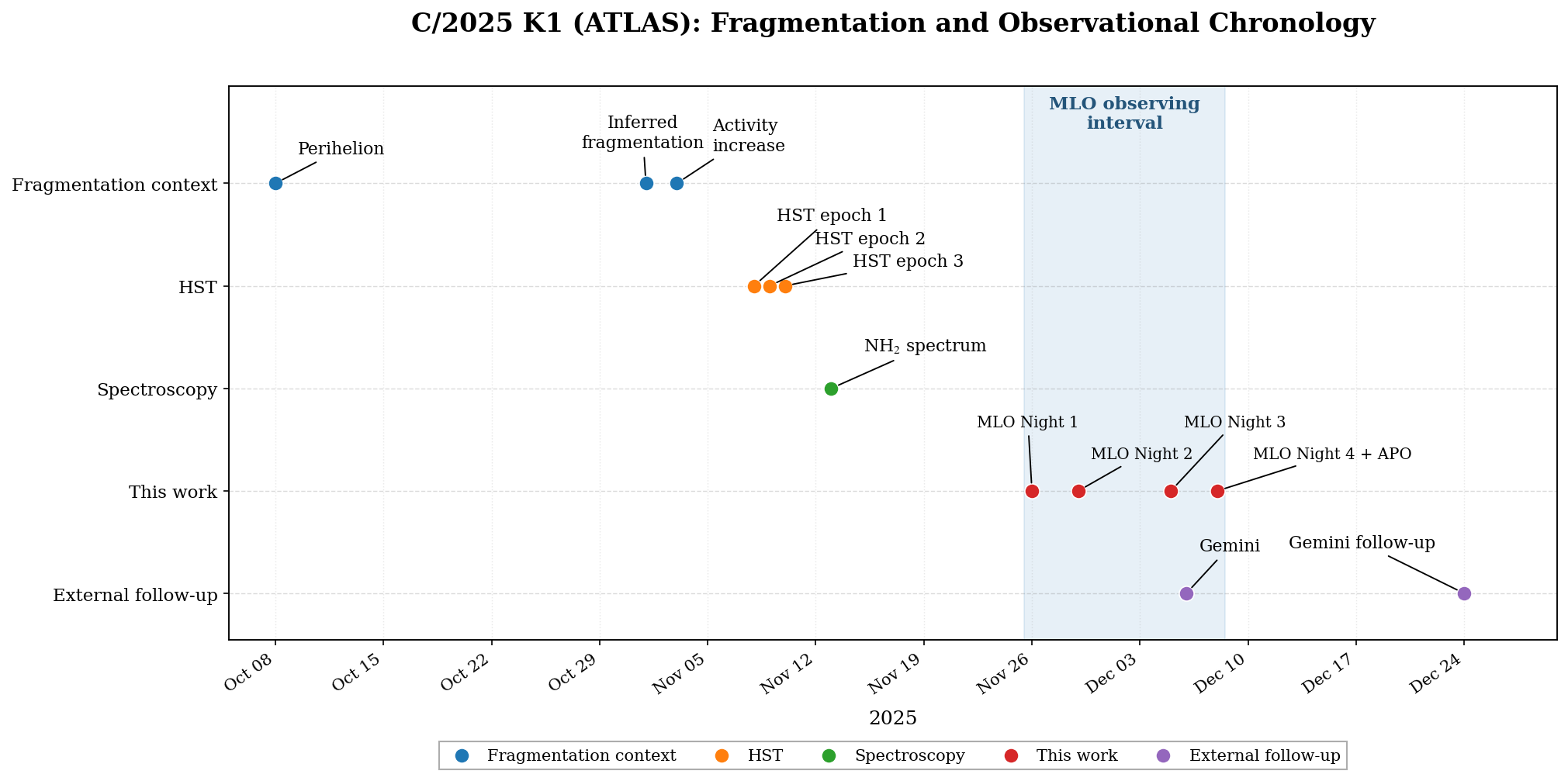}
\caption{Fragmentation and observational chronology for C/2025 K1 (ATLAS). The shaded interval marks the four MLO epochs analyzed in this work. Fragmentation context and HST epochs are from \citet{Bodewits2026}; the November 13 spectrum is from \citet{Ganesh2025}; and the later Gemini/APO epochs are from \citet{Ingebretsen2026}.}
\label{fig:timeline}
\end{figure*}

Here we analyze four Mount Laguna Observatory (MLO) epochs obtained on 2025 November 26, November 29, December 5, and December 8, bridging the interval between the early HST fragmentation observations and the later Gemini/APO campaign (Figure~\ref{fig:timeline}). By the beginning of the MLO sequence, more than two weeks had elapsed since HST resolved the breakup. Our observations therefore do not probe the initial fragmentation event itself, but instead follow the subsequent evolution of an unresolved fragment-coma complex. We quantify the morphology of the inner coma along the projected fragment axis, measure calibrated $B-V$ and $V-R$ colors, convert those colors to solar-normalized broadband reflectivity gradients, and test whether temporal changes survive PSF homogenization, profile-extraction systematics, stellar contamination, photometric calibration uncertainty, aperture choice, and changing physical aperture scale. The central goal is to determine whether independently measured morphological and broadband spectral-shape evolution persists after the initial breakup and, if so, what constraints can be placed on the continued restructuring of the fragmented inner coma.

\section{Observations and Data Reduction}\label{sec:obs}

\subsection{MLO imaging}\label{subsec:observations}

The observations were obtained with the Mount Laguna Observatory 1.0-m (40-inch) reflector using a 2048 × 2048-pixel E2V CCD42-40 detector with a native plate scale of 0.358 arcsec pixel\(^{-1}\). The detector was operated with 2 × 2 binning to better match the typical 1–2 arcsec seeing at the observatory, giving a nominal image scale of approximately 0.716 arcsec per binned pixel. This agrees closely with the approximately 0.718 arcsec pixel\(^{-1}\) scale measured from our astrometric solutions. The recorded FITS images have a raw format of $1100\times1024$ pixels including the overscan region, with the illuminated science region occupying $1024\times1024$ binned pixels. Detector temperatures during the observations were near $-110^\circ$C.

The final analysis uses Johnson--Cousins $B$, $V$, and $R$ imaging. The raw K1 sequences contain 27 exposures on November 26 ($B$: 6, $V$: 6, $R$: 9, $I$: 6), 86 on November 29 ($B$: 21, $V$: 21, $R$: 23, $I$: 21), 18 on December 5 ($B$: 5, $V$: 5, $R$: 8), and 28 on December 8 ($B$: 6, $V$: 6, $R$: 10, $I$: 6). Exposure times were 30 s on November 26 and 15 s on the remaining nights. Although $I$-band observations were obtained on three of the four nights, the final precision analysis was restricted to $BVR$ because the November 26 $I$ frames show a strong fixed-pattern fringe and no suitable $I$-band flat-field sequence was available for December 8.

The final $BVR$ inventory contains 125 accepted frames. On December 8 one $V$ exposure, a105, produced a repeatable ASTAP plate scale of approximately 0.715768 arcsec pixel$^{-1}$ compared with approximately 0.71784 arcsec pixel$^{-1}$ for the other frames. Its field center was consistent with neighboring images, but the anomalous scale persisted after a fresh independent astrometric solution. The frame was therefore excluded from astrometry-dependent morphology and from the final accepted-inventory photometry. The accepted December 8 $BVR$ inventory is thus 6 $B$, 5 $V$, and 10 $R$ frames.

\begin{deluxetable*}{lrrrrrrr}
\tablecaption{MLO observing circumstances for C/2025 K1 (ATLAS)\label{tab:obs}}
\tablehead{
\colhead{UT Date} &
\colhead{$r_h$ (au)} &
\colhead{$\Delta$ (au)} &
\colhead{$\alpha$ (deg)} &
\colhead{$N_B$} &
\colhead{$N_V$} &
\colhead{$N_R$} &
\colhead{$t_{\rm exp}$ (s)}
}
\startdata
2025 Nov 26 & 1.2268 & 0.4053 & 45.65 & 6 & 6 & 9 & 30 \\
2025 Nov 29 & 1.2826 & 0.4301 & 39.03 & 21 & 21 & 23 & 15 \\
2025 Dec 5 & 1.3920 & 0.5335 & 32.62 & 5 & 5 & 8 & 15 \\
2025 Dec 8 & 1.4455 & 0.6037 & 31.78 & 6 & 5 & 10 & 15 \\
\enddata
\tablecomments{$r_h$ is the heliocentric distance, $\Delta$ is the observer-comet distance, and $\alpha$ is the solar phase angle. The frame counts are the $BVR$ exposures entering the final analysis. The December 8 $V$ count excludes exposure a105.}
\end{deluxetable*}

\subsection{CCD calibration and astrometry}\label{subsec:ccd}

The raw detector contains 1024 illuminated columns followed by a short transition region and a stable overscan. The reduction measured the row-by-row overscan from columns 1030--1099, subtracted that value from each row, and trimmed the data to the active $1024\times1024$ region. Master biases were formed by median combination. On November 26, 25 bias frames were combined to form the master bias, and the final twilight master flats combined 9 $B$, 9 $V$, 6 $R$, and 5 $I$ exposures; only the $BVR$ products are used here. No calibration sequence was present in the supplied November 29 data set. Those frames were therefore reduced using the November 26 overscan-corrected master bias and filter-specific master flats. Subsequent stellar photometric calibration was performed independently on each science exposure, reducing sensitivity to night-to-night photometric zero-point differences, although any residual spatial flat-field mismatch remains a potential systematic limitation. December 5 used 25 same-night biases and five dome flats per $B$, $V$, and $R$ filter. December 8 used 28 same-night biases and five twilight flats per $B$, $V$, and $R$ filter.

Astrometric solutions were obtained with the ASTAP command-line solver using the D80 reference database and an explicit 0.20 deg field of view. The solutions were written into calibrated FITS WCS headers and validated through the resulting plate scale and field center. Representative accepted scales are approximately 0.718 arcsec pixel$^{-1}$. Ephemerides were obtained with JPL Horizons \citep{Giorgini1996} at exposure midpoints for the primary K1 component and Horizons-listed B, C, and D components, using unique Horizons record identifiers to avoid designation ambiguity. The MLO observer location used in Horizons was longitude $-116.4256^\circ$, latitude $32.8394^\circ$, and elevation 1.859 km.

\subsection{Stellar photometric calibration}\label{subsec:calibration}

Photometric calibration used the Gaia DR3 Synthetic Photometry Catalogue (GSPC), including standardized Johnson--Kron--Cousins $B$, $V$, and $R_C$ magnitudes derived from Gaia BP/RP spectrophotometry \citep{GaiaSynthetic2023}. Stellar instrumental magnitudes were measured with a 12-pixel circular aperture (approximately 8.6 arcsec) and a local sky annulus from 18 to 28 pixels. Candidate standards were required to have signal-to-noise ratio above 30, no close catalog neighbor within 35 pixels, separation of at least 120 pixels from the comet center, and local peak values below 50,000 ADU. A loose 0.50 mag frame-median cut removed catastrophic stellar measurements before the final calibration fit.

Instrumental magnitudes were defined as
\begin{equation}
m_{\rm inst}=-2.5\log_{10}\left(\frac{F_{\rm net}}{t_{\rm exp}}\right),
\end{equation}
where $F_{\rm net}$ is the sky-subtracted aperture flux and $t_{\rm exp}$ is the exposure time.

For each filter we simultaneously fitted one zero point per exposure and one global color coefficient,
\begin{equation}
m_{\rm cat}-m_{\rm inst}=ZP_{\rm frame}+C_f({\rm color}-{\rm pivot}),
\end{equation}
using weighted least squares, a 0.005 mag uncertainty floor, and iterative 3$\sigma$ robust clipping. The adopted transformations are
\begin{align}
B &= b_{\rm inst}+Z_B+0.0834[(B-V)-0.7408],\\
V &= v_{\rm inst}+Z_V-0.0918[(B-V)-0.7379],\\
R_C &= r_{\rm inst}+Z_R+0.0301[(V-R)-0.4617].
\end{align}
The final fit retains 259/275 measurements in $B$, 397/417 in $V$, and 741/811 in $R_C$, with RMS residuals of 0.0241, 0.0247, and 0.0237 mag, respectively (Table~\ref{tab:cal}).

\begin{deluxetable}{lrrrrr}
\tablecaption{Global photometric transformation fits\label{tab:cal}}
\tablehead{\colhead{Band} & \colhead{Color} & \colhead{Pivot} & \colhead{$C_f$} & \colhead{$\sigma(C_f)$} & \colhead{RMS (mag)}}
\startdata
$B$ & $B-V$ & 0.7408 & +0.08338 & 0.00882 & 0.0241\\
$V$ & $B-V$ & 0.7379 & -0.09183 & 0.00671 & 0.0247\\
$R_C$ & $V-R$ & 0.4617 & +0.03013 & 0.00798 & 0.0237\\
\enddata
\end{deluxetable}

The November 29 $B$ exposure a127 had a zero point approximately 0.215 mag below the nightly median but retained seven accepted standards and a small residual scatter, so it was kept and corrected by its frame-specific zero point. December 5 $B$ was a different case: the five frames retained only one or two standards each. We therefore replaced their individual zero points with a pooled nightly value of 21.40756 mag and propagated a common 0.01169 mag zero-point uncertainty rather than allowing that uncertainty to average down across frames.

\section{Analysis}\label{sec:analysis}

\subsection{Common-PSF morphology}\label{subsec:morphmethod}

The morphology analysis was performed only after homogenizing the $BVR$ observations to common two-dimensional effective PSFs. Because variations in seeing and PSF ellipticity between filters and observing nights can broaden or displace low-contrast inner-coma structure, temporal morphology comparisons were made only between images matched to the same effective resolution.

The effective PSF was measured independently on every accepted $BVR$ science frame using field stars well separated from the comet. Candidate stellar sources were identified after mild Gaussian smoothing and were required to exceed $8\sigma$ above the global background. Sources within 100 pixels of the predicted comet position were excluded. For each accepted source, a local background was estimated in an annulus spanning 10--13 pixels from the stellar center, and the two-dimensional second moments of the positive stellar signal within a 10-pixel radius were used to construct a $2\times2$ covariance matrix. Sources with signal-to-noise ratios below 30 or anomalous widths or elongations were rejected. This procedure produced 4735 accepted stellar PSF measurements across all 125 accepted $BVR$ frames.

For each science frame, the effective stellar covariance was obtained from the median covariance elements of the accepted stars. An equal-frame-weight effective covariance for each night and filter was then obtained from the median of the corresponding frame-level covariance elements. The resulting effective PSFs had major-axis FWHM values ranging from approximately 3.45 to 5.15 arcsec, with the largest measured value of 5.153 arcsec occurring in the November 26 $B$ data.

The primary morphology analysis adopted a circular Gaussian target PSF with FWHM 5.40 arcsec, slightly broader than the largest measured input PSF. For each night/filter combination, the covariance of the required convolution kernel was calculated as

\begin{equation}
\mathbf{C}_{\rm ker}
=
\mathbf{C}_{\rm target}
-
\mathbf{C}_{\rm obs},
\end{equation}

where $\mathbf{C}_{\rm obs}$ is the measured two-dimensional effective-PSF covariance and $\mathbf{C}_{\rm target}$ is the covariance of the circular target PSF. All 12 resulting kernel covariance matrices were positive definite. Each covariance difference was represented by a normalized two-dimensional Gaussian convolution kernel extending to $5\sigma$ along its broadest dimension. Convolution was performed with a NaN-aware FFT procedure in which the image and valid-pixel weight map were convolved separately and subsequently normalized.

PSF matching was applied to the nightly/filter comet-registered stacks. The complete morphology analysis was also repeated using common circular target PSFs of 5.3, 5.4, 5.6, and 6.0 arcsec to test sensitivity to the adopted common resolution. The 5.40 arcsec products are used for the nominal presentation, while temporal conclusions were required to remain consistent across the tested target PSFs.

\begin{figure*}
\centering
\includegraphics[width=0.86\textwidth]{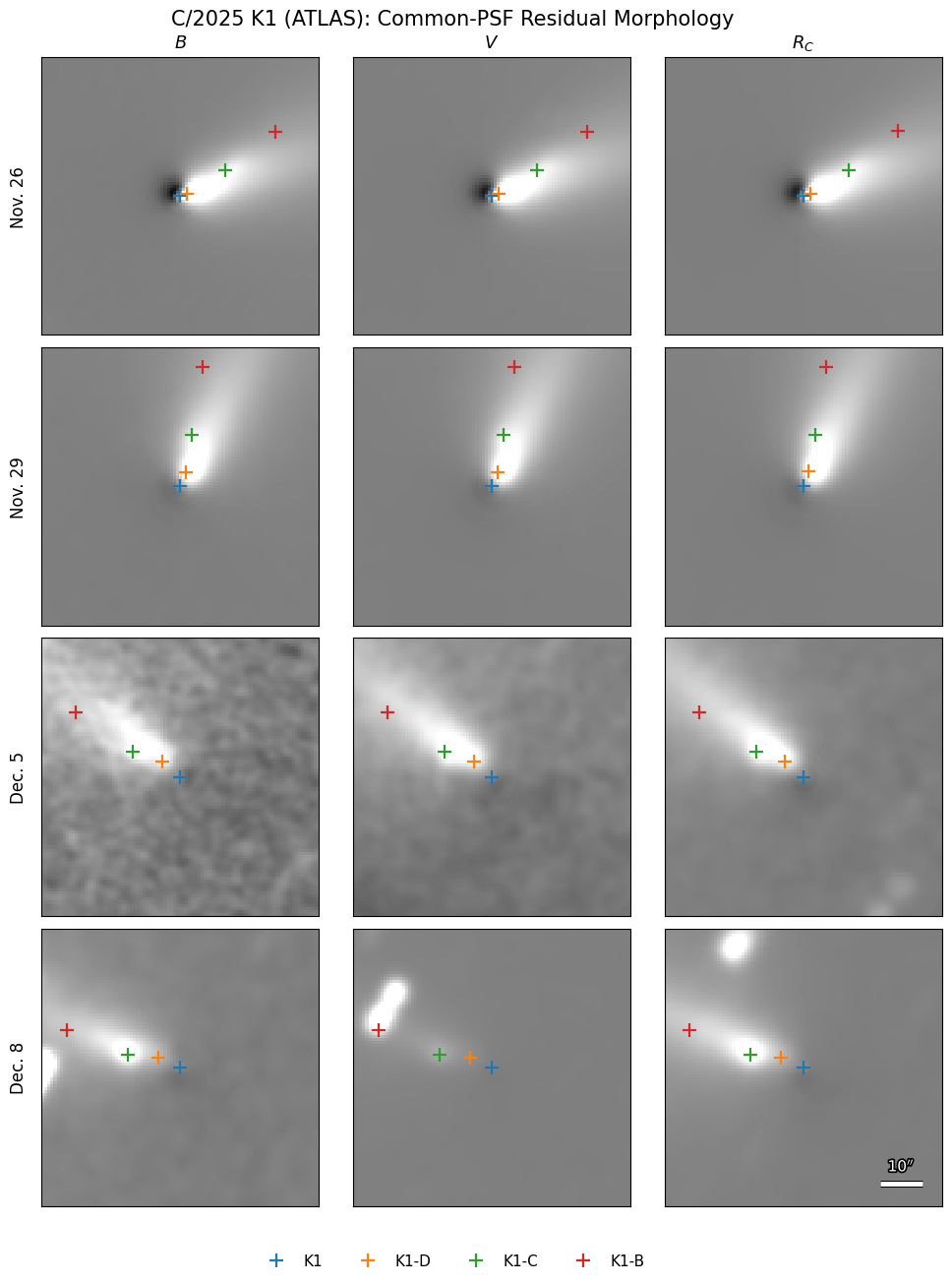}
\caption{Common-5.4-arcsec-PSF azimuthal-median coma-subtracted residual morphology in $B$, $V$, and $R_C$ for the four observing epochs. Colored crosses mark the predicted positions of K1, K1-D, K1-C, and K1-B, as identified by the legend. All panels have been homogenized to the same circular effective PSF before comparison. Each panel is independently displayed with a symmetric intensity scale chosen to emphasize morphology; displayed surface brightnesses should therefore not be compared quantitatively between panels. The images illustrate extended asymmetric structure along the projected fragment axis, while quantitative enhancement positions are obtained from the profile analysis described in Section~\ref{subsec:morphmethod}.}
\label{fig:morphpanels}
\end{figure*}

Following PSF matching, an azimuthal-median coma model centered on K1 was constructed for each night/filter stack and subtracted to suppress the approximately symmetric component of the coma and emphasize asymmetric residual structure along the projected fragment axis. The resulting common-PSF residual images are shown in Figure~\ref{fig:morphpanels}.

The position of the localized residual enhancement was quantified using one-dimensional profiles extracted along the projected K1-to-B fragment axis. For the nominal extraction, the profile was formed using a strip half-width of 4 pixels. A quadratic local baseline was fitted over projected distances of 5.5--19 arcsec from K1 while excluding the 8.5--15.5 arcsec region containing the enhancement. After subtraction of this local baseline, the enhancement position was defined as the maximum residual within a search interval of 8--16 arcsec.

To determine whether the inferred enhancement position depended on a particular profile-extraction choice, the analysis was repeated over a grid of 324 combinations of strip half-width, polynomial degree, baseline-fit interval, excluded interval, and peak-search interval. Configurations for which the selected maximum occurred at a search-window boundary were rejected because such solutions do not provide a well-localized interior peak. This criterion left 216 valid broadband configurations on November 26 and all 324 configurations on November 29, December 5, and December 8. Temporal changes were subsequently evaluated using matched extraction configurations between epochs, so that each comparison used the same methodological choices on both nights.

The measured structure is extended at the spatial resolution of the MLO observations and is therefore not interpreted as a resolved fragment nucleus. Throughout this work we refer to it as a \emph{fragment-axis surface-brightness enhancement}. When describing its location relative to the Horizons fragment predictions, we use the phrase \emph{surface-brightness enhancement spatially associated with the projected K1-C region}. The measured enhancement position traces the morphology of the unresolved fragment-coma complex and is not treated as direct astrometry or orbital motion of fragment C.

\subsection{Comet photometry and contamination control}\label{subsec:cometphot}

Comet growth curves were measured at radii of 5, 7.5, 10, 12.5, 15, 20, 25, and 30 arcsec with a sky annulus from 45 to 65 arcsec. During the four epochs the predicted K1-D separation is approximately 4--6 arcsec, K1-C lies near 12.2--12.7 arcsec, and K1-B lies near 28 arcsec. A 7.5 arcsec radius was adopted as the primary color aperture: it improves signal-to-noise over 5 arcsec while remaining interior to the nominal C location and is less vulnerable to field-source contamination than 10 arcsec and larger apertures.

A strict Gaia-neighbor rejection applied to individual comet apertures removed too much of the December 5 data. The final method instead uses photometrically calibrated, comet-registered cubes with no stellar mask in the central comet aperture and takes the pixelwise temporal median. Field stars move relative to the comet registration and are therefore strongly suppressed by the median, while Gaia masks are retained in the sky annulus. Where fully Gaia-star-masked stacks provide adequate central coverage, the two methods agree with RMS differences of 0.00311 mag in $B-V$ and 0.00339 mag in $V-R$.

\subsection{Photometric uncertainty model}\label{subsec:uncertainty}

Science-frame sampling uncertainty was estimated with 1000 bootstrap realizations for each night/filter combination. In each realization, the accepted comet-registered calibrated frames were resampled with replacement using the original number of frames, and a new pixelwise temporal-median stack was constructed. Photometry was then repeated in the 7.5 arcsec aperture using the 45--65 arcsec sky annulus. All 12 night/filter combinations produced 1000 valid bootstrap measurements.

The final color Monte Carlo additionally propagates the photometric zero-point and transformation-coefficient uncertainties. For night/filter sequences with independent frame zero points, the effective zero-point uncertainty of the median stack was estimated by drawing each frame zero point from its fitted uncertainty distribution and taking the median. For the pooled December 5 $B$ calibration, the adopted zero-point error was instead treated as a single common nightly uncertainty.

One draw of each global transformation coefficient was shared across all nights within a realization because these coefficients were fitted globally across the data set. The calibration draws were combined with the science-frame bootstrap distributions before solving for $B-V$ and $V-R$, thereby preserving the dominant calibration covariance in temporal color differences.

\subsection{Solar-normalized broadband reflectance}\label{subsec:reflectance}

We adopt the model-independent solar colors $(B-V)_\odot=0.653\pm0.003$ and $(V-R)_\odot=0.356\pm0.003$ from \citet{Ramirez2012}. The analysis uses representative Johnson--Cousins effective wavelengths of 4361, 5448, and 6407 \AAng\ for $B$, $V$, and $R_C$, respectively \citep{Bessell2005}. With reflectance normalized to unity in $V$,
\begin{align}
R_B/R_V &= 10^{-0.4[(B-V)_c-(B-V)_\odot]},\\
R_R/R_V &= 10^{+0.4[(V-R)_c-(V-R)_\odot]}.
\end{align}
For two bands,
\begin{equation}
\spr = \frac{200000}{\lambda_2-\lambda_1}\,
\frac{R_2-R_1}{R_2+R_1},
\end{equation}
with $\lambda$ in \AAng\ and $\spr$ in percent per 1000 \AAng. We report $\spr_{B-V}$, $\spr_{V-R}$, and the full-baseline $\spr_{B-R}$. We also define the three-point broadband curvature index
\begin{equation}
C_{\rm BRV}\equiv\spr_{V-R}-\spr_{B-V}.
\end{equation}
This is a diagnostic of broadband spectral shape, not a spectroscopic curvature measurement.

\section{Results}\label{sec:results}

\subsection{Fragment-axis morphology evolution}\label{subsec:morphresults}

The final common-PSF method distributions yield median enhancement positions of $13.498^{+1.053}_{-0.862}$ arcsec on November 26, $11.679^{+0.048}_{-0.120}$ arcsec on November 29, $11.749^{+0.072}_{-0.048}$ arcsec on December 5, and $12.754^{+0.096}_{-0.167}$ arcsec on December 8, where the intervals are the 16th--84th percentiles across valid extraction methods (Table~\ref{tab:morph}). The November 26 distribution is substantially broader than the later distributions.

Temporal changes were evaluated configuration by configuration using the same extraction method at both epochs. November 26 to 29 gives a mean change of $-1.970\pm0.804$ arcsec across 216 matched valid configurations, spanning $-3.303$ to $-0.981$ arcsec; every configuration is negative. November 29 to December 5 gives $+0.109\pm0.119$ arcsec and is method sensitive, with a range from $-0.050$ to $+0.333$ arcsec. December 5 to 8 gives $+0.966\pm0.162$ arcsec, spanning $+0.717$ to $+1.196$ arcsec; every configuration is positive. The negative November 26--29 sign and positive December 5--8 sign are also recovered at every tested common target PSF from 5.3 to 6.0 arcsec.

Relative to the predicted K1-C location, the mean matched changes are $-2.241$ arcsec, $+0.029$ arcsec, and $+0.889$ arcsec for the three consecutive intervals. These values trace the movement of an extended brightness enhancement relative to the projected fragment geometry and are not direct fragment-C motions.

\begin{deluxetable}{lrrrr}
\tablecaption{Common-PSF fragment-axis enhancement\label{tab:morph}}
\tablehead{\colhead{UT Date} & \colhead{$N_{\rm valid}$} & \colhead{Median peak} & \colhead{16--84\% peak} & \colhead{Median peak$-C$}\\
& & \colhead{(arcsec)} & \colhead{(arcsec)} & \colhead{(arcsec)}}
\startdata
2025 Nov 26 & 216 & 13.498 & 12.637--14.551 & +1.276\\
2025 Nov 29 & 324 & 11.679 & 11.560--11.727 & -0.814\\
2025 Dec 5  & 324 & 11.749 & 11.701--11.821 & -0.825\\
2025 Dec 8  & 324 & 12.754 & 12.586--12.849 & +0.103\\
\enddata
\tablecomments{The distribution is across valid broadband extraction configurations after boundary-hit configurations are removed.}
\end{deluxetable}

\begin{figure}
\centering
\includegraphics[width=\columnwidth]{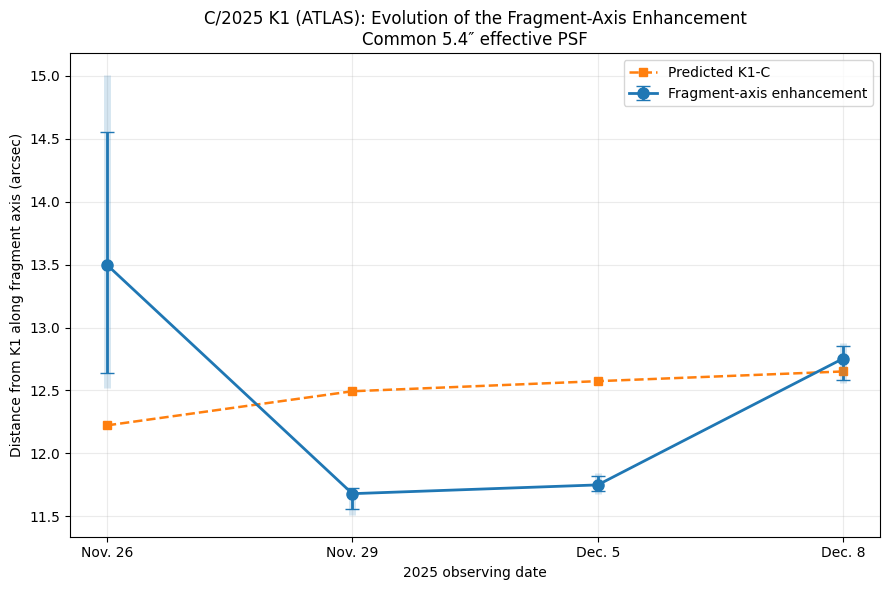}
\caption{Temporal evolution of the fragment-axis surface-brightness enhancement in the nominal common-5.4-arcsec-PSF analysis. Points show the median enhancement position across valid extraction configurations, with intervals spanning the 16th--84th percentiles. The dashed sequence shows the predicted K1-C separation from JPL Horizons. The enhancement positions trace extended coma morphology and are not interpreted as fragment-C astrometry.}
\label{fig:morphevolution}
\end{figure}

\subsection{Inner-coma $BVR$ colors}\label{subsec:colorresults}

\begin{figure*}
\centering
\includegraphics[width=0.96\textwidth]{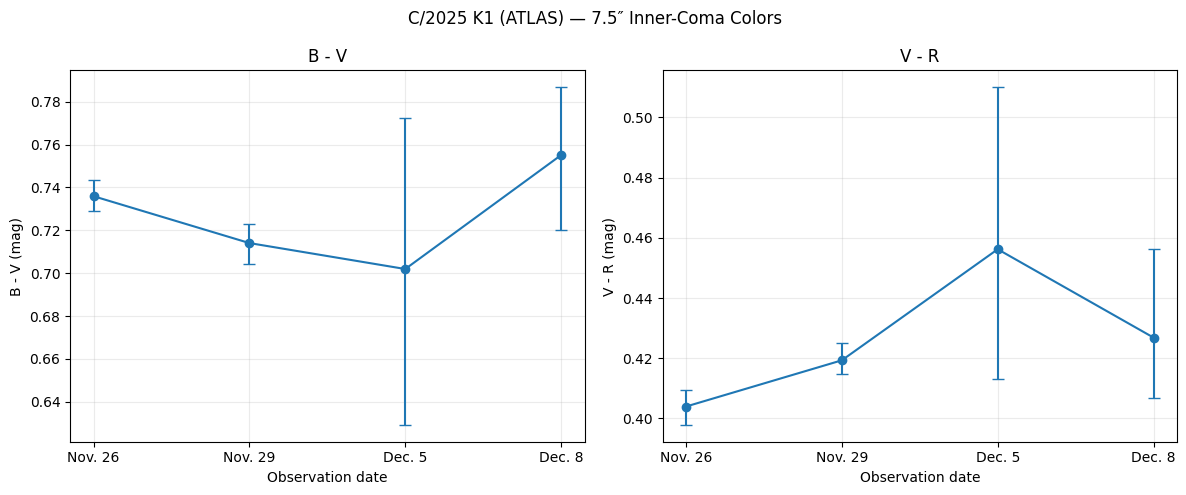}
\caption{Evolution of the 7.5 arcsec inner-coma $B-V$ and $V-R$ colors. Error bars show the 16th--84th percentile intervals after propagation of science-frame bootstrap, nightly zero-point, and global color-term uncertainties.}
\label{fig:colors}
\end{figure*}

The final 7.5 arcsec colors are shown in Figure~\ref{fig:colors} and summarized in Table~\ref{tab:photresults}. November 26 gives $B-V=0.7359^{+0.0077}_{-0.0068}$ and $V-R=0.4040^{+0.0054}_{-0.0061}$ mag. On November 29 the corresponding values are $0.7141^{+0.0087}_{-0.0098}$ and $0.4193^{+0.0058}_{-0.0045}$ mag.

The November 26--29 Monte Carlo differences are
\begin{align}
\Delta(B-V) &= -0.02296^{+0.01257}_{-0.01219}\ {\rm mag},\\
\Delta(V-R) &= +0.01614^{+0.00769}_{-0.00792}\ {\rm mag}.
\end{align}
The $B-V$ change is negative in 96.9\% of realizations and the $V-R$ change is positive in 98.4\%. Their signs are also invariant over the independently measured 5, 7.5, and 10 arcsec apertures, although the amplitudes are aperture dependent (Appendix~\ref{app:photo}). We therefore describe the November 26--29 color evolution as repeatable and suggestive rather than as an aperture-independent high-significance amplitude measurement.

The December 5 and December 8 color distributions are much broader. Neither the November 29--December 5 nor the December 5--8 color difference has a 16th--84th percentile interval that remains one-sided in both color indices, so we do not claim significant later color evolution.

\subsection{Solar-normalized broadband reflectivity}\label{subsec:sloperesults}

All nominal epochs are redder than the adopted solar colors over the full $B$--$R$ baseline. The best-constrained values are $\spr_{B-R}=5.90^{+0.32}_{-0.32}$\% per 1000 \AAng\ on November 26 and $5.60^{+0.41}_{-0.41}$\% per 1000 \AAng\ on November 29. Their difference,
\begin{equation}
\Delta\spr_{B-R}=-0.32^{+0.47}_{-0.44}\ {\rm \%~per~1000~\AAng},
\end{equation}
crosses zero, with 74.8\% of realizations negative. Thus the total visible redness changes little.

\begin{figure}
\centering
\includegraphics[width=\columnwidth]{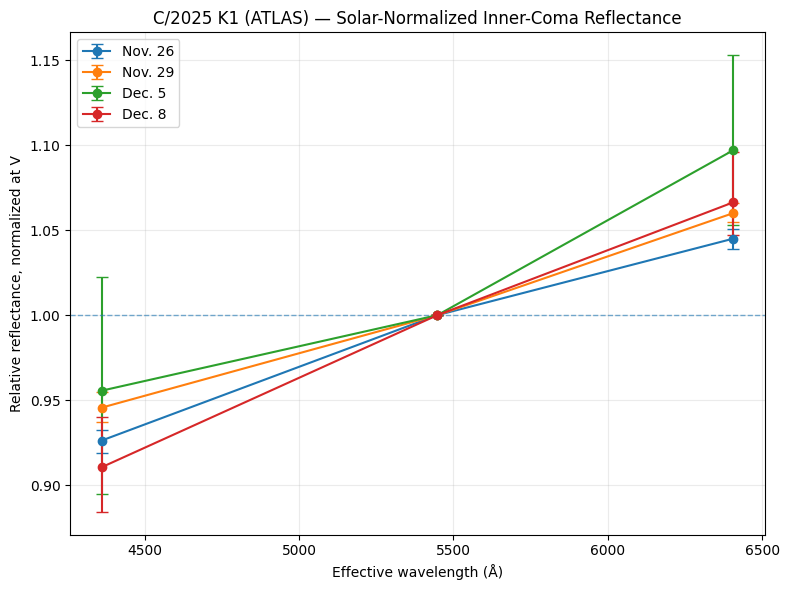}
\caption{Solar-normalized three-point $BVR$ reflectance, normalized at $V$. These are broadband measurements and should not be interpreted as gas-free dust spectra.}
\label{fig:reflectance}
\end{figure}

\begin{figure*}
\centering
\includegraphics[width=0.94\textwidth]{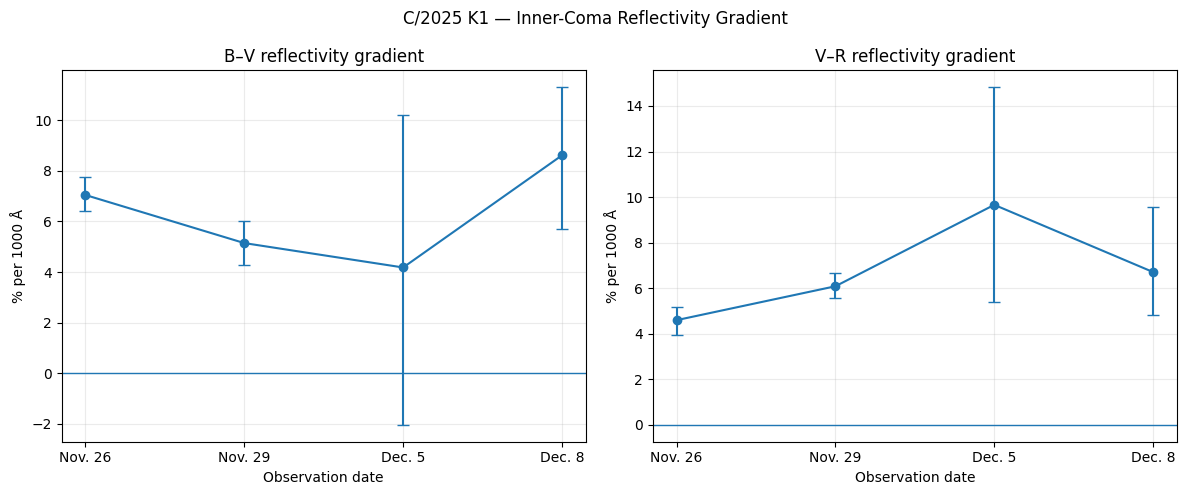}
\caption{Evolution of the $B-V$ and $V-R$ normalized reflectivity gradients. The opposite November 26--29 changes motivate the broadband-curvature diagnostic.}
\label{fig:gradients}
\end{figure*}

The segmented slopes evolve in opposite directions. $\spr_{B-V}$ decreases from 7.05 to 5.14\% per 1000 \AAng, while $\spr_{V-R}$ increases from 4.60 to 6.08\% per 1000 \AAng. Figures~\ref{fig:reflectance} and \ref{fig:gradients} show the three-point solar-normalized reflectance and the segmented slope evolution.

\subsection{Broadband spectral-shape evolution}\label{subsec:curvresults}

The curvature index changes from $-2.45^{+1.05}_{-1.10}$\% per 1000 \AAng\ on November 26 to $+0.93^{+1.21}_{-1.17}$\% per 1000 \AAng\ on November 29. The temporal difference is
\begin{equation}
\Delta C_{\rm BRV}=+3.47^{+1.50}_{-1.65}\ {\rm \%~per~1000~\AAng},
\end{equation}
with 99.1\% of Monte Carlo realizations positive. The later curvature changes are poorly constrained and their central 68\% intervals cross zero.

\begin{figure}
\centering
\includegraphics[width=\columnwidth]{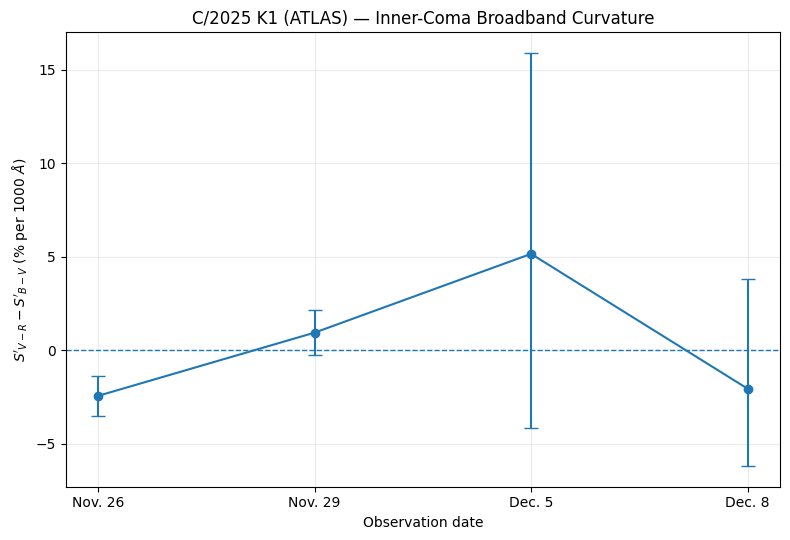}
\caption{Broadband curvature index $C_{\rm BRV}=\spr_{V-R}-\spr_{B-V}$. The best-supported temporal change is the positive November 26--29 shift; the later epochs are much less precise.}
\label{fig:curvature}
\end{figure}

\begin{deluxetable*}{lrrrr}
\tablecaption{Final 7.5 arcsec broadband measurements\label{tab:photresults}}
\tablehead{\colhead{UT Date} & \colhead{$B-V$} & \colhead{$V-R$} & \colhead{$\spr_{B-R}$} & \colhead{$C_{\rm BRV}$}\\
& \colhead{(mag)} & \colhead{(mag)} & \colhead{(\%/1000 \AAng)} & \colhead{(\%/1000 \AAng)}}
\startdata
2025 Nov 26 & $0.7359^{+0.0077}_{-0.0068}$ & $0.4040^{+0.0054}_{-0.0061}$ & $5.90^{+0.32}_{-0.32}$ & $-2.45^{+1.05}_{-1.10}$\\
2025 Nov 29 & $0.7141^{+0.0087}_{-0.0098}$ & $0.4193^{+0.0058}_{-0.0045}$ & $5.60^{+0.41}_{-0.41}$ & $+0.93^{+1.21}_{-1.17}$\\
2025 Dec 5  & $0.7019^{+0.0704}_{-0.0728}$ & $0.4562^{+0.0538}_{-0.0433}$ & $6.73^{+2.31}_{-2.05}$ & $+5.15^{+10.75}_{-9.32}$\\
2025 Dec 8  & $0.7551^{+0.0317}_{-0.0350}$ & $0.4269^{+0.0294}_{-0.0202}$ & $7.72^{+1.10}_{-0.79}$ & $-2.06^{+5.86}_{-4.14}$\\
\enddata
\tablecomments{Uncertainties are 16th--84th percentile Monte Carlo intervals.}
\end{deluxetable*}

\subsection{Fixed-physical-aperture sensitivity test}\label{subsec:physicalap}

The 7.5 arcsec aperture corresponds to projected radii of 2205, 2339, 2902, and 3284 km across the four epochs because the observer-comet distance increased from 0.405 to 0.604 au. To test the key November comparison, we fixed the projected radius at the November 26 value of 2205 km. The required angular radius is 7.50 arcsec on November 26 and 7.07 arcsec on November 29. Linear interpolation within the independently measured 5--7.5--10 arcsec color grid gives
\begin{align}
\Delta(B-V)_{2205\,{\rm km}} &= -0.02050\ {\rm mag},\\
\Delta(V-R)_{2205\,{\rm km}} &= +0.01382\ {\rm mag},
\end{align}
compared with $-0.02188$ and $+0.01441$ mag from the nominal fixed 7.5 arcsec temporal-median colors. The changes differ by only 0.00138 and 0.00059 mag, respectively. The November 26--29 result is therefore not driven by the changing geocentric distance. We use the corresponding December interpolations only as sensitivity checks because those epochs have stronger aperture dependence and larger photometric uncertainties.

\section{Discussion}\label{sec:discussion}

\subsection{Evolution after the major fragmentation}\label{subsec:postfrag}

Our first MLO epoch occurs more than two weeks after HST resolved the fragmentation sequence of C/2025 K1 \citep{Bodewits2026}. The observations presented here therefore do not probe the initiation of the breakup itself. Instead, they sample the subsequent evolution of an unresolved fragment-coma complex in which multiple components, their individual comae, and dust released during and after fragmentation contribute simultaneously within the MLO resolution element.

Such complexity is characteristic of well-observed split comets. Fragmentation can produce multiple active bodies whose relative brightnesses evolve independently and whose surrounding comae overlap in projection \citep{Boehnhardt2004}. Comet 73P/Schwassmann--Wachmann 3 provides a particularly relevant example: its principal components displayed substantially different and time-variable coma morphologies, while one component underwent repeated secondary fragmentation and developed transient fragments and asymmetric coma structures \citep{Vincent2010}. The disruption of C/1999 S4 (LINEAR) likewise produced numerous fragments embedded within a large dust complex after the major breakup \citep{Weaver2001}. These examples illustrate that post-fragmentation morphology can evolve through changing activity and fragment weighting even when individual nuclei cannot be cleanly isolated.

The morphology measured for C/2025 K1 is similarly non-monotonic. The fragment-axis surface-brightness enhancement relocates inward between November 26 and 29, remains approximately stable through December 5, and then relocates outward by December 8. The first and last changes retain the same sign across all valid matched extraction configurations and across every tested common target PSF from 5.3 to 6.0 arcsec. The intermediate November 29--December 5 interval is much smaller and method sensitive.

Because the enhancement is spatially extended, this behavior should not be interpreted as the trajectory of a single nucleus or as evidence that fragment C physically reversed direction. Instead, the measured position represents the location of a maximum in the projected surface-brightness distribution along the fragment axis. Changes in the relative brightnesses of unresolved fragments, overlapping fragment comae, anisotropic dust production, secondary fragmentation, dust expansion, and changing viewing geometry can all shift this maximum without requiring comparable motion of an individual solid body. The Horizons fragment positions therefore serve only as geometric references for the evolving morphology.

\subsection{Coincident morphology and broadband spectral-shape evolution}\label{subsec:coincident}

The strongest broadband spectral-shape change occurs during the same November 26--29 interval as the strongest morphology relocation. This coincidence is noteworthy because the two measurements are methodologically independent. The morphology result is derived from common-PSF residual images and survives hundreds of profile-extraction choices, whereas the color result is obtained from independently calibrated temporal-median photometry with science-frame and calibration uncertainties propagated through a separate Monte Carlo analysis.

Between November 26 and 29, $B-V$ decreases while $V-R$ increases. The respective directions occur in 96.9\% and 98.4\% of the final Monte Carlo realizations and remain the same at aperture radii of 5, 7.5, and 10 arcsec. Repeating the comparison at a fixed projected radius of 2205 km produces nearly the same changes, demonstrating that the result is not produced by the increasing observer-comet distance. At the same time, every valid matched morphology configuration places the November 29 enhancement inward of its November 26 location.

The temporal coincidence does not by itself establish a causal connection between the morphological and photometric changes. It is nevertheless consistent with a common evolution of the unresolved fragment-coma complex. For example, a change in the relative contribution of one fragment and its surrounding coma could shift the location of the dominant surface-brightness enhancement while simultaneously changing the mixture of dust and molecular emission included within the photometric aperture. Continued secondary fragmentation or changing fragment activity could produce a similar effect. The present observations do not uniquely distinguish among these possibilities.

Importantly, the November color evolution is not well described as a simple global blueing or reddening. The full $B$--$R$ reflectivity gradient changes only from approximately 5.90 to 5.60\% per 1000~\AAng, with a temporal difference whose central uncertainty interval crosses zero. Instead, the segmented slopes evolve in opposite directions: $\spr_{B-V}$ decreases while $\spr_{V-R}$ increases. The observational result is therefore better described as a redistribution of broadband spectral slope around $V$ than as a substantial change in the total visible redness of the coma.

\subsection{Optical-color context}\label{subsec:colorcontext}

The positive full-baseline reflectivity gradients measured at all four epochs indicate an integrated optical spectrum that is redder than the adopted solar colors. This behavior is broadly consistent with the optical reddening commonly measured in cometary dust comae. Cometary dust continua are generally red relative to the solar spectrum, although individual comets exhibit substantial dispersion in color and spectral slope \citep{JewittMeech1986}. Larger population studies likewise show broadly similar mean optical dust colors among short-period and long-period comets despite their different dynamical source regions \citep{Jewitt2015}.

In that context, the overall $B$--$R$ redness of C/2025 K1 is not by itself the unusual aspect of the November measurements. More significant for the present analysis is that the total redness remains approximately stable while the relative $B-V$ and $V-R$ contributions change in opposite directions. The resulting positive shift in $C_{\rm BRV}=\spr_{V-R}-\spr_{B-V}$ is recovered in 99.1\% of the Monte Carlo realizations. This three-band diagnostic is useful for quantifying the change, but it should not be interpreted as a measurement of spectroscopic continuum curvature. With only three broadband measurements, several physically different spectral-energy distributions can generate the same value of $C_{\rm BRV}$.

The absence of a large change in $\spr_{B-R}$ also argues against interpreting the November 26--29 event as a wholesale change in the optical properties of all material within the aperture. A changing mixture of components provides a less restrictive interpretation: individual fragments, dust populations, or gaseous emissions with different spectral characteristics may change their relative contributions while leaving the integrated $B$--$R$ slope nearly unchanged.

\subsection{Dust and gas contributions}\label{subsec:dustgas}

The solar-normalized slopes reported here are broadband reflectivity diagnostics and are not guaranteed to represent a pure dust continuum. This distinction is especially important for C/2025 K1. Spectroscopy obtained on November 13, less than two weeks before the first MLO epoch, was dominated by NH$_2$ emission while CN, C$_2$, and C$_3$ were reported as weak \citep{Ganesh2025}. Molecular emission falling within the broad Johnson--Cousins passbands could therefore contribute to the observed colors and to their temporal evolution.

Fragmentation also provides several mechanisms by which the dust contribution itself could evolve. Newly exposed material can alter the local production of grains, while changes in the grain-size distribution, dust outflow, and the relative activities of individual fragments can modify the integrated scattering properties within a fixed aperture. Fragmenting comets can remain highly variable well after the initial breakup, as illustrated by the repeated secondary fragmentation and changing coma structures observed in 73P/Schwassmann--Wachmann 3 \citep{Vincent2010}. The MLO data alone cannot determine whether the November spectral-shape change is dominated by changing dust properties, molecular emission, changing fragment weights, or a combination of these effects.

For this reason, our strongest photometric conclusion is deliberately phenomenological: the integrated inner coma undergoes a repeatable November 26--29 change in broadband spectral shape while its overall $B$--$R$ redness remains approximately stable. Contemporaneous spectroscopy, spectrophotometry, or narrowband gas and continuum imaging would be required to assign this evolution to a specific dust or molecular component.

\subsection{Relation to the Gemini/APO observations}\label{subsec:gemini}

The later Gemini North and ARC/APO observations provide an important independent view of the system near the end of the MLO sequence. \citet{Ingebretsen2026} resolve at least four fragments in observations obtained between December 6 and 24 and report an unusually blue Sloan $g-r$ color near 0.40. They also find color differences among the comet and resolved fragments and discuss particle-size and light-scattering effects as possible contributors to those differences.

The near-contemporaneous December 8 observations are particularly useful qualitatively. The resolved Gemini/APO measurements demonstrate that C/2025 K1 remained a multiply fragmented system at the end of our observing interval, whereas the MLO measurements combine much of that structure within a substantially broader effective PSF. This supports our treatment of the MLO photometry and morphology as measurements of an integrated fragment-coma complex rather than of a single nucleus.

A direct numerical comparison of the colors is not justified, however. The Gemini/APO measurements use Sloan-like passbands and substantially higher spatial resolution, and individual fragments can be separated in ways that are impossible in the MLO data. The adopted apertures and relative weighting of fragment and coma light are also different. In addition, the fragment labels used by \citet{Ingebretsen2026} are not assumed here to map one-to-one onto the Horizons B/C/D component labels used as geometric references in our analysis. Synthetic photometry based on contemporaneous spectra, or observations obtained simultaneously in both photometric systems, would be required for a physically meaningful direct comparison.

The unusually blue later measurements nevertheless reinforce the broader conclusion that the post-fragmentation optical properties of the system were evolving and potentially heterogeneous. Our December 5 and December 8 Johnson--Cousins colors are not precise enough to establish a significant continuation of the November color evolution, so we do not attempt to connect the two data sets through a single temporal color trend.

\subsection{Physical interpretation and limitations}\label{subsec:limitations}

Taken together, the observations are consistent with continued restructuring of the unresolved inner coma after the major early-November fragmentation. HST had already demonstrated that C/2025 K1 underwent hierarchical fragmentation \citep{Bodewits2026}, and the later Gemini/APO observations show that multiple components persisted into December \citep{Ingebretsen2026}. The MLO sequence fills part of the interval between those resolved observations and shows that the projected distribution of optical surface brightness continued to change during this period.

Our measurements do not identify the physical mechanism responsible for the original fragmentation. Processes such as rotational instability, thermal stress, internal gas pressure, and other structural failures can contribute to cometary disruption, and outgassing-driven rotational instability has been proposed as an important destruction mechanism for some small long-period comets \citep{Jewitt2022}. Because our observations begin weeks after the inferred primary breakup and do not resolve the individual nuclei, they cannot discriminate among such mechanisms for C/2025 K1. Nor can we determine whether the later morphological changes require additional fragmentation rather than changing activity among fragments that already existed.

Several observational limitations also affect the interpretation. First, the MLO spatial resolution is insufficient to isolate clean point-source nuclei for the individual fragments. The measured fragment-axis enhancement therefore cannot be converted into fragment astrometry. Second, the three broadband filters cannot uniquely separate dust scattering from molecular emission, particularly given the unusual NH$_2$-dominated spectrum reported shortly after fragmentation. Third, photometric precision varies between epochs. December 5 has particularly weak $B$-band standard-star leverage, while the December 8 colors are also substantially less precise than the November measurements.

The November 29 images additionally lack a same-night calibration sequence and were reduced using the November 26 master bias and filter-specific flats. Independent frame-by-frame stellar zero points correct photometric throughput differences but cannot guarantee removal of every possible residual spatial flat-field mismatch. We therefore regard this as an additional systematic limitation. The November 26--29 color result is strengthened, however, by the large number of accepted stellar calibration measurements, the consistency of the color-change direction across independent aperture radii, the temporal-median versus Gaia-masked validation, and the nearly identical result obtained when the projected physical aperture is held fixed.

Finally, a fixed angular aperture encompasses an increasing physical radius as the comet recedes from Earth. The explicit fixed-physical-aperture test shows that this effect does not explain the key November 26--29 color change, but it becomes increasingly important for the later epochs. These limitations motivate our emphasis on robust directions of change, cross-method consistency, and the temporal coincidence of independently measured morphology and broadband spectral-shape evolution rather than on a unique compositional or dynamical interpretation.

Within those constraints, the MLO observations provide evidence that the aftermath of the major fragmentation was not static. The spatial distribution of optical brightness continued to reorganize over timescales of days, and the best-constrained interval of morphological evolution coincided with a measurable redistribution of the integrated broadband spectral shape. Higher-resolution imaging combined with simultaneous spectroscopy would be required to determine whether such behavior is driven primarily by secondary fragmentation, changing fragment activity, dust evolution, molecular emission, or some combination of these processes.

\section{Conclusions}\label{sec:conclusions}

We present four epochs of post-fragmentation $BVR$ imaging of C/2025 K1 (ATLAS) from 2025 November 26 through December 8. Our principal conclusions are:

\begin{enumerate}
\item After common two-dimensional PSF matching, the fragment-axis surface-brightness enhancement relocates robustly inward from November 26 to 29. Every valid configuration in the 324-method grid gives the same sign, and the sign remains negative at common target PSFs from 5.3 to 6.0 arcsec.

\item The November 29--December 5 morphology is approximately stable and method sensitive. From December 5 to 8 the enhancement relocates robustly outward, again with every valid configuration giving the same sign.

\item The 7.5 arcsec inner-coma colors change over November 26--29, with $B-V$ decreasing and $V-R$ increasing. The respective directions occur in 96.9\% and 98.4\% of the final Monte Carlo realizations and persist at 5, 7.5, and 10 arcsec aperture radii.

\item The full $B-R$ reflectivity gradient remains approximately constant over the same interval, while $C_{\rm BRV}=\spr_{V-R}-\spr_{B-V}$ changes by $+3.47^{+1.50}_{-1.65}$\% per 1000 \AAng. The curvature change is positive in 99.1\% of realizations, so the photometric result is better described as evolution in broadband spectral shape than as a large change in total optical redness.

\item Repeating the key color comparison at a fixed projected radius of 2205 km produces nearly identical changes, showing that the November 26--29 result is not an artifact of the changing geocentric distance.
\end{enumerate}

The temporal coincidence between the strongest morphology relocation and the best-supported broadband spectral-shape change is consistent with continued restructuring of the unresolved fragmented inner coma. Higher-resolution imaging and contemporaneous spectroscopy are needed to distinguish changing fragment weights, dust evolution, and molecular emission.

\begin{acknowledgments}
The author acknowledges the use of the 1.0-m (40-inch) reflector at Mount Laguna Observatory. The author thanks Robert Quimby for obtaining the observations on 2025 November 26 and November 29 and for coordinating the observing program. The author also thanks Alexander Nielsen and Paige Keating for obtaining the observations on 2025 December 5, Hy Truong and Katarzyna Mccoy for obtaining the observations on 2025 December 8, and John ``Chuck'' Horst for guiding the student observers during the December observing runs. The author is grateful to the Mount Laguna Observatory and San Diego State University personnel who supported the observations and data acquisition. This research made use of the Gaia DR3 Synthetic Photometry Catalogue and the VizieR catalogue access tool, CDS, Strasbourg, France. Ephemerides were obtained using the JPL Horizons On-Line Ephemeris System.

\end{acknowledgments}

\facility{MLO:1m}

\software{ASTAP, Astropy \citep{Astropy2013}, Astroquery, Matplotlib, NumPy, pandas, SciPy}

\section*{Data Availability}

The raw CCD observations used in this work are publicly available through Zenodo \citep{Walker2026Data}. The archive contains the original science frames and available calibration frames obtained with the Mount Laguna Observatory 1.0-m (40-inch) reflector on 2025 November 26, November 29, December 5, and December 8. No same-night calibration sequence was available for the November 29 observations; as described in Section~\ref{subsec:ccd}, those data were reduced using the November 26 master calibration products.

\section*{AI Disclosure}

ChatGPT (OpenAI) was used under the author's supervision to assist with code development and debugging. All data products, numerical results, references, figures, and scientific interpretations included in the final manuscript were verified by the author, who assumes full responsibility for the work.

\appendix

\section{Color-aperture robustness}\label{app:photo}

The temporal-median colors were also measured at 5 and 10 arcsec. For November 26--29, $\Delta(B-V)$ is $-0.00990$, $-0.02188$, and $-0.03559$ mag at 5, 7.5, and 10 arcsec, respectively, while $\Delta(V-R)$ is $+0.01221$, $+0.01441$, and $+0.02417$ mag. The sign of both changes is therefore invariant across the tested aperture range, although the amplitude depends on aperture.

\begin{figure*}
\centering
\includegraphics[width=0.92\textwidth]{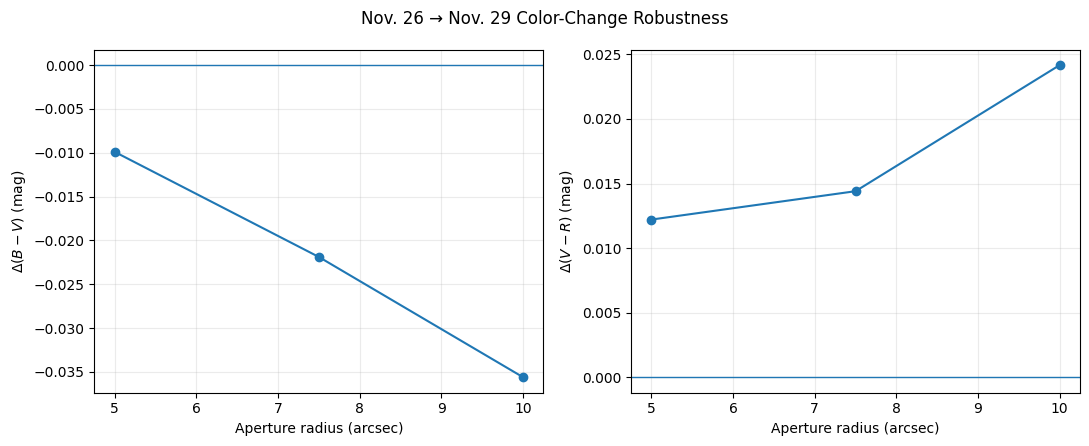}
\caption{Aperture dependence of the November 26--29 color change. The direction is consistent from 5 to 10 arcsec, while the amplitude changes with aperture.}
\label{fig:aperturerobust}
\end{figure*}

\section{Morphology extraction diagnostics}\label{app:morph}

\begin{figure*}
\centering
\includegraphics[width=0.96\textwidth]{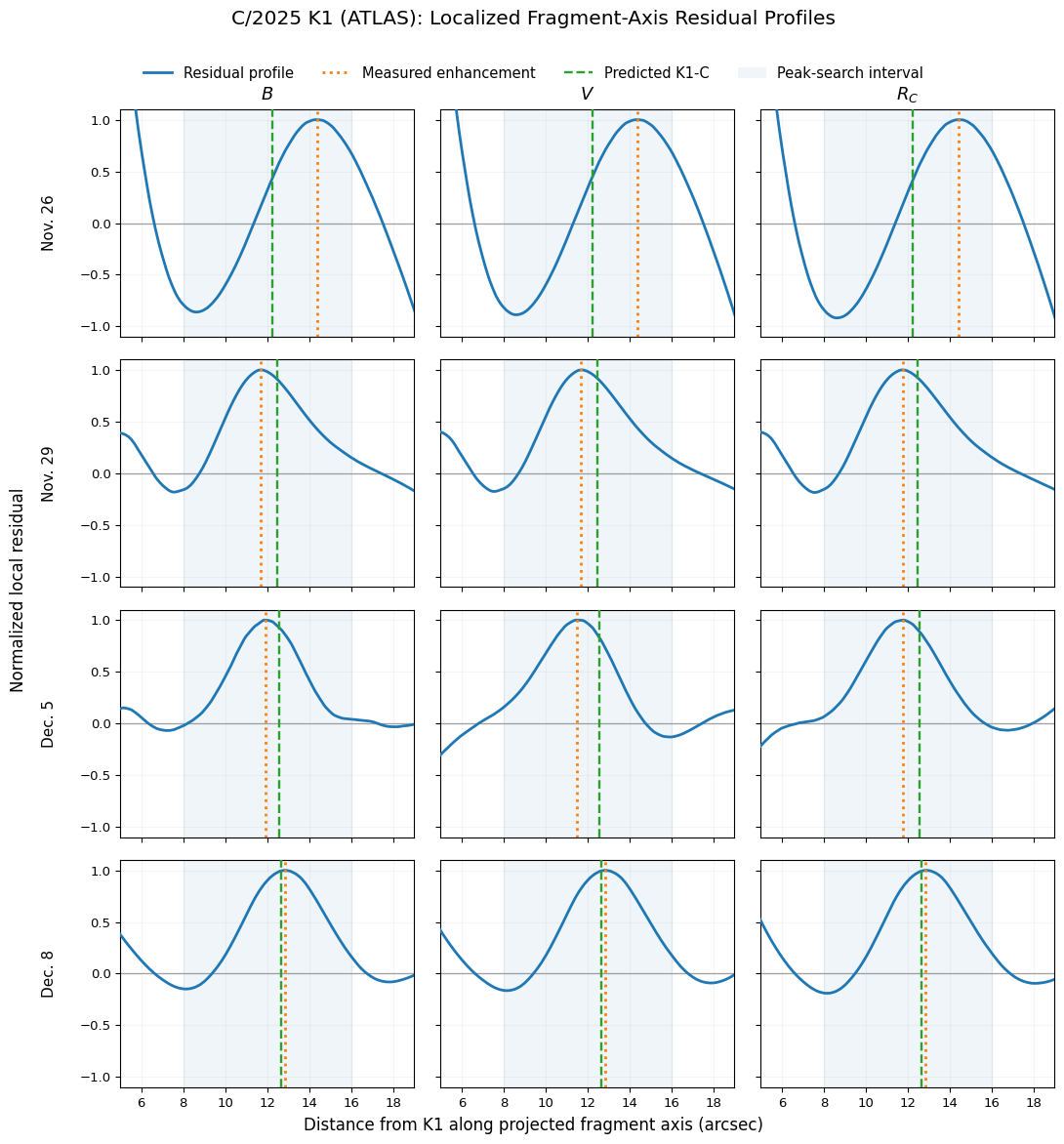}
\caption{Nominal localized fragment-axis residual profiles for the four observing epochs and three filters after common-PSF processing and subtraction of the local baseline. Each profile is normalized independently for visualization, so residual amplitudes should not be compared quantitatively between panels. Dotted vertical lines mark the measured surface-brightness enhancement, dashed vertical lines show the predicted K1-C separation, and the shaded region indicates the nominal 8--16 arcsec peak-search interval. The enhancement remains spatially extended; these profiles are used to localize a surface-brightness feature rather than a resolved nucleus.}
\label{fig:localprofiles}
\end{figure*}

Figure~\ref{fig:localprofiles} shows the nominal localized residual profiles after common-PSF matching and local baseline subtraction. The vertical markers compare the measured enhancement and predicted K1-C location. The systematic grid was designed to test whether conclusions depend on the precise strip width, polynomial degree, baseline window, exclusion window, or peak-search interval.

Across matched extraction configurations, the November 26--29 broadband change has mean $-1.9698$ arcsec, standard deviation 0.8039 arcsec, and range $-3.3028$ to $-0.9813$ arcsec. The December 5--8 change has mean $+0.9658$ arcsec, standard deviation 0.1621 arcsec, and range $+0.7174$ to $+1.1960$ arcsec. The intermediate November 29--December 5 interval has mean $+0.1089\pm0.1190$ arcsec and is sign sensitive.

\bibliographystyle{aasjournalv7.1}
\bibliography{references}

\end{document}